\documentclass[final,5p,times,twocolumn]{elsarticle}
\usepackage{lineno,hyperref,amsmath,amsthm,amssymb,amsfonts,ragged2e,color,subfig}
\journal{``Physics of Plasmas"}
\begin{document}
\begin{frontmatter}
\title{Dust-ion-acoustic rogue waves in presence of non-extensive non-thermal electrons}
\author{T. I. Rajib$^{*1}$, N. K. Tamanna$^{**,1}$, N. A. Chowdhury$^{\dag,2}$, A. Mannan$^{\ddag,1,3}$, S. Sultana$^{\S,1,4}$, and A. A. Mamun$^{\S\S,1}$}
\address{$^1$Department of Physics, Jahangirnagar University, Savar, Dhaka-1342, Bangladesh\\
$^2$Plasma Physics Division, Atomic Energy Centre, Dhaka-1000, Bangladesh\\
$^3$Institut f{\"u}r Mathematik, Martin Luther Universit{\"a}t Halle-Wittenberg, D-06099 Halle, Germany\\
$^4$Lehrstuhl Theoretische Physik IV, Ruhr Universit{\"a}t Bochum, 44780 Bochum, Germany\\
e-mail: $^*$tirajibphys@gmail.com, $^{**}$tamanna1995phy@gmail.com, $^{\dag}$nurealam1743phy@gmail.com,\\ 
$^\ddag$abdulmannan@juniv.edu, $^\S$ssultana@juniv.edu,  $^{\S\S}$mamun\_phys@juniv.edu,}
\begin{abstract}
Dust-ion-acoustic (DIA) rogue waves (DIARWs) are investigated in a three components dusty plasma system
containing inertialess electrons featuring non-thermal non-extensive distribution as well as
inertial warm ions and negative dust grains. A nonlinear Schr\"{o}dinger equation (NLSE), which governs
the conditions of the modulational instability (MI) of DIA waves (DIAWs), is obtained by using
the reductive perturbation method. It has been observed from the numerical analysis of NLSE that the plasma system supports both
modulationally stable domain in which dispersive and nonlinear coefficients of the NLSE have same
sign and unstable domain in which dispersive and nonlinear coefficients of the NLSE have opposite
sign, and also supports the DIARWs only in the unstable domain. It is also observed  that the basic features (viz.
stability of the DIAWs, MI, growth rate, amplitude, and width of the DIARWs, etc.) are
significantly modified by the related plasma parameters (viz. dust charge state, number density of electron and ion, non-extensive
parameter q, and non-thermal parameter $\alpha$, etc.). The present study is useful for understanding the mechanism of the formation of
DIARWs in the laboratory and space environments where inertialess mixed
distributed  electrons can exist with inertial ions and dust grains.
\end{abstract}
\end{frontmatter}
\section{Introduction}
\label{1sec:Introduction}
The investigation of plasma system containing massive dust grains has a great importance
due to its wide applications in laboratory and space dusty plasma
medium (DPM) \cite{Shukla1992,Barkan1996,Rapp2005,Kim2006,Mirsa2009,Duha2011,Sahu2012}.
Dust grains can be either negatively/positively charged massive object, if
they are injected into any plasma system \cite{Rapp2005,Kim2006}, depending on the
ambient environments. Such kind of highly charged and extremely massive dust grains can modify the normal plasma modes. Among the modified modes,
the dust-ion-acoustic (DIA) waves (DIAWs), which theoretically proposed by Shukla and Silin \cite{Shukla1992}
and experimentally confirmed by Barkan \textit{et al.} \cite{Barkan1996}, has received much attention.
Significant research efforts have been devoted in investigating the characteristics of DIAWs in the pair-ion DPM (PIDPM) \cite{Mirsa2009}, and degenerate DPM \cite{Mirsa2009,Duha2011}.
Misra \cite{Mirsa2009} studied the nonlinear propagation of DIA shocks waves in a quantum PIDPM, and
numerically discussed  the stationary  shock solutions. Duha \textit{et al.} \cite{Duha2011}
considered a Multi-ion DPM, and observed both the solitary and shocks structures. Sahu and Tribeche \cite{Sahu2012}
reported the electrostatic dust-acoustic (DA) solitary and shock structures in DPM having non-thermal plasma species.

The nonlinear features of plasma waves are rigorously affected by the velocity distribution function
of the plasma particles, and the Maxwellian velocity distribution is one of the most commonly used velocity
distribution functions in plasma physics.  The space observations \cite{Young2005} and laboratory experiments \cite{Lundin1989}
indicated the presence of particles which deviates from the Maxwellian distribution. These findings
lead to the modification of the Boltzmann-Gibbs-Shannon entropy which has been recognized by Renyi \cite{Renyi1955}
and successively proposed by Tsallis \cite{Tsallis1988}. The index $q$ in the non-extensive $q$-distribution
underpins the generalized Tsallis entropy, and is connected to the dynamics of the long range interacting systems, and also measures
the degree of its non-extensivity of the plasma species, and  has been successfully applied
to explain the intrigue  mechanism of a number of complex plasma situations \cite{Emamuddin2013,Bacha2012}.
On the other hand, Cairns \textit{et al.} \cite{Cairns1995} introduced
a distribution to describe the dynamics of the highly energetic tails, in which a parameter $\alpha$
can measure the deviation of the plasma species from the Maxwellian velocity distribution function, in
space plasma environments \cite{Paul2016,Tasnim2014}.
Paul and Bandyopadhyay \cite{Paul2016} investigated DIA solitary waves in a multi-component DPM having
non-thermal electrons. Tasnim \textit{et al.} \cite{Tasnim2014} examined the effects of non-thermal plasma species
on the the formation of DA Gardner solitons in DPM. A number of authors have considered a  hybrid Cairn-Tsallis/non-thermal
non-extensive distribution \cite{Tribeche2012,Wang2013,Guo2014,Guo2015,Williams2013,El-Depsy2016,Amour2012,Bouzit2015} for analyzing the
 nonlinear properties of the plasma medium and found that this hybrid Cairn-Tsallis/non-thermal
non-extensive  distribution is suitable to provide a better picture which coincides with a wide range of the space
observational data \cite{Taibanya2019,Ait2012}.

Rogue wave (RW) \cite{Akhmediev2009,Kedziora2011,Bains2013}, which governs by the
rational solution of the standard nonlinear Schr\"{o}dinger equation
(NLSE) \cite{Akhmediev2009,Kedziora2011,Bains2013,Kourakis2003,Fedele2002,Sultana2011,Rahman2018,El-Labany2015,Hassan2019,Misra2006}
in a dispersive medium, was firstly observed in the ocean. It is a rare, short-lived, and
high-energy event with amplitude much higher than the average wave crests around it \cite{Akhmediev2009}.
The usual feature of a RW is that it will appear suddenly and increase up to a very
high amplitude with the exponential growth, and finally disappear without a trace \cite{Kedziora2011}.
Recently, a number of authors have employed the reductive perturbation method (RPM) to establish the NLSE for the
investigation of modulational instability (MI) of the DA waves (DAWs) and also the formation of DA RWs (DARWs)
in various plasma system \cite{Rahman2018,El-Labany2015,Hassan2019,Misra2006}.
Bains \textit{et al.} \cite{Bains2013} theoretically and numerically analyzed the
effect of non-extensivity of electrons and ions on the MI criteria of DAWs.
Rahman \textit{et al.} \cite{Rahman2018} investigated the MI conditions of the DAWs in presence of the non-thermal plasma species,
and observed that the maximum value of the MI growth rate increases with the non-extensive parameter $q$, and also found that
the amplitude and width of the RWs increase with an increase in the value of the non-thermal parameter $\alpha$.
El-Labany \textit{et al.} \cite{El-Labany2015} have considered a three components DPM having non-thermal plasma
species for investigation of the DARWs, and found that the nonlinearity of the DPM increases with $\alpha$.
Hassan \textit{et al.} \cite{Hassan2019} analyzed theoretically and numerically the instability
conditions of the ion-acoustic waves in a non-thermal plasma, and
highlighted that the non-thermality of the electrons enhances the nonlinearity of the
plasma medium as well as the height and thickness of the RWs.
Misra and Chowdhury \cite{Misra2006} considered a three components plasma model having non-thermal Cairn's distributed ions, and
observed that the envelope solitons are significantly modified by the presence of non-thermal ions.

In DAWs, the moment of inertia is provided by the dust grains and restoring force is provided by the thermal pressure of the
ions and electrons. On the other hand in DIAWs, the moment of inertia is provided by the ions and restoring force
is provided by the thermal pressure of the electrons in presence of immobile dust grains. The mass and charge of the dust grains
are considerably larger than the ions while the mass and charge of the ion are considerably larger
than the electron. It may be noted here that in the DIAWs, if anyone consider
the pressure term of the ions then it is important to be considered the moment of inertia of the ions along with
the dust grains in presence of inertialess electrons. This means that the consideration of the pressure term
of the ions highly contributes to the moment of inertia along with inertial dust grains to generate DIAWs in a DPM
having inertialess electrons. In the present work, we are interested to investigate the nonlinear propagation of DIARWs in which the moment
of inertia is provided by the inertial warm ions and negatively charged dust grains and the restoring force is
provided by the thermal pressure of the inertialess electrons in a three components DPM (dusts-ions-electrons)
by using the standard NLSE.

The layout of the manuscript is as follows: The governing equations describing our plasma
model are presented in Section \ref{1sec:Governing Equations}. The derivation of a NLSE
is given in Section \ref{1sec:Derivation of the NLSE}. The MI and RWs are discussed in
Section \ref{1sec:Modulational Instability and Rogue Waves}. Numerical analysis
is presented in Section \ref{1sec:Numerical analysis}. A  conclusion is provided in Section \ref{1sec:Conclusion}.
\section{Governing Equations}
\label{1sec:Governing Equations}
We consider a three components DPM comprising of inertial positively charged warm ion (charge $q_i=Z_ie$
and mass $m_i$) and inertial negatively charged dust grains (charge $q_d=-Z_de$ and mass $m_d$) as
well as inertialess non-thermal non-extensive electron (charge $q_e=-e$; mass $m_e$);
where $Z_i$ ($Z_d$) is the number of  protons (electrons) residing onto the ion (dust grains)
surface, and $e$ is the magnitude of the charge of an electron.
Overall, the charge neutrality condition for our plasma model can
be written as $ Z_in_{i0} = Z_d n_{d0}+ n_{e0}$.  Now, the normalized governing equations
of the DIAWs can be written as
\begin{eqnarray}
&&\hspace*{-1.3cm}\frac{\partial n_d}{\partial t}+\frac{\partial}{\partial x}(n_d u_d)=0,
\label{1eq:1}\\
&&\hspace*{-1.3cm}\frac{\partial u_d}{\partial t} + u_d\frac{\partial u_d }{\partial x}=\rho_1\frac{\partial \phi}{\partial x},
\label{1eq:2}\\
&&\hspace*{-1.3cm}\frac{\partial n_i}{\partial t}+\frac{\partial}{\partial x}(n_i u_i)=0,
\label{1eq:3}\\
&&\hspace*{-1.3cm}\frac{\partial u_i}{\partial t} + u_i\frac{\partial u_i}{\partial x}+\rho_2n_i\frac{\partial n_i}{\partial x}= -\frac{\partial \phi}{\partial x},
\label{1eq:4}\\
&&\hspace*{-1.3cm}\frac{\partial^2\phi}{\partial x^2}+n_i= \rho_3n_e+(1-\rho_3)n_d,
\label{1eq:5}\
\end{eqnarray}
where $n_d$ $(n_i)$ is the dust (ion) number density normalized by its equilibrium
value $n_{d0} $ $(n_{i0})$; $u_d$ $(u_i)$ is the dust (ion) fluid speed normalized by
the ion-acoustic wave speed $C_i=(Z_i k_BT_e/m_i)^{1/2}$ with $T_e$ being the non-thermal non-extensive
electron temperature, and $k_B$ being the Boltzmann constant; $\phi$ is the electrostatic wave potential normalized
by $k_BT_e/e$; the time and space variables are normalized by ${\omega^{-1}_{pi}}=(m_i/4\pi {Z_i}^2 e^2 n_{i0})^{1/2}$
and $\lambda_{Di}=(k_BT_e/4 \pi Z_i e^2 n_{i0})^{1/2}$, respectively. The pressure term of the ion is recognized as
$P_i=P_{i0}(N_i/n_{i0})^\gamma$ with $P_{i0}=n_{i0}k_BT_i$ being the equilibrium
pressure of the ion, and $T_i$ being the temperature of warm ion, and
$\gamma=(N+2)/N$ (where $N$ is the degree of freedom and for one-dimensional case
$N=1$, then $\gamma=3$). Other plasma parameters are
$\rho_1=\rho\mu$, $\rho=Z_d/Z_i$, $\mu=m_i/m_d$, $\rho_2=3T_{i}/Z_iT_{e}$, and $\rho_3=n_{e0}/Z_in_{i0}$, etc.
Now, the expression for the number density of non-thermal non-extensive electrons following the
non-thermal non-extensive distribution \cite{Tribeche2012,Wang2013,Guo2014,Guo2015,Williams2013} can be written as
\begin{eqnarray}
&&\hspace*{-1.3cm}n_e=[1+A\phi+B\phi^2]\times[1+(q-1)\phi]^{\frac{(q+1)}{2(q-1)}},
\label{1eq:6}\
\end{eqnarray}
where the parameter $q$ stands for the strength of non-extensive system and the
coefficients $A$  and $B$ are defined by $A=-16q\alpha/(3-14q+15q^2+12\alpha)$
and $B=-A(2q-1)$. Here $\alpha$ is a parameter determining the number of
non-thermal electrons in the model. Williams \textit{et al.} \cite{Williams2013} discussed the
range and the validity of ($q$, $\alpha$) for solitons. In the limiting case $(q\rightarrow1$ and $\alpha= 0$),
the above distribution reduces to the well-known Maxwell-Boltzmann velocity distribution.
For ($q\rightarrow1$ and  $\alpha\neq 0$), the above distribution reduces to Cairn distribution.

Now, by substituting Eq. \eqref{1eq:6} into Eq. \eqref{1eq:5}, and expanding up to
third order of $\phi$, we get
\begin{eqnarray}
&&\hspace*{-1.3cm}\frac{\partial^2\phi}{\partial x^2}+ n_{i}= \rho_3+(1-\rho_3)n_d+S_1\phi
\nonumber\\
&&\hspace*{+0.3cm}+S_2\phi^2+S_3\phi^3+\cdot\cdot\cdot,
\label{1eq:7}\
\end{eqnarray}
where
\begin{eqnarray}
&&\hspace*{-1.3cm}S_1=[2\rho_3A+\rho_3(q+1)]/2,
\nonumber\\
&&\hspace*{-1.3cm}S_2=[8\rho_3B+4\rho_3A(q+1)-\rho_3(q+1)(q-3)]/8,
\nonumber\\
&&\hspace*{-1.3cm}S_3=[24\rho_3B(q+1)-6\rho_3A(q+1)(q-3)+M_1]/48,
\nonumber\
\end{eqnarray}
with $M_1=\rho_3(q+1)(q-3)(3q-5)$. The terms containing $S_1$, $S_2$, and $S_3$ in Eq. \eqref{1eq:7}
are due to the contribution of the non-thermal non-extensive electrons.
\section{Derivation of the NLSE}
\label{1sec:Derivation of the NLSE}
To study the MI of the DIAWs, we want to derive the NLSE by employing the RPM
and for that case, first we can write the stretched co-ordinates in the form
\begin{eqnarray}
&&\hspace*{-1.3cm}\xi={\epsilon}(x-v_g t),
\label{1eq:8}\\
&&\hspace*{-1.3cm}\tau={\epsilon}^2 t,
\label{1eq:9}\
\end{eqnarray}
where $v_g$ is the group speed and $\epsilon$ $(0<\epsilon<1)$ is a small parameter measuring
the weakness of the dispersion. Then we can write the dependent variables as
\begin{eqnarray}
&&\hspace*{-1.3cm}n_{d}=1+\sum_{m=1}^{\infty}\epsilon^{m}\sum_{l=-\infty}^{\infty}n_{dl}^{(m)}(\xi,\tau)~\mbox{e}^{i l(kx-\omega t)},
\label{1eq:10}\\
&&\hspace*{-1.3cm}u_{d}=\sum_{m=1}^{\infty}\epsilon^{m}\sum_{l=-\infty}^{\infty}u_{dl}^{(m)}(\xi,\tau)~\mbox{e}^{i l(kx-\omega t)},
\label{1eq:11}\\
&&\hspace*{-1.3cm}n_{i}=1+\sum_{m=1}^{\infty}\epsilon^{m}\sum_{l=-\infty}^{\infty}n_{il}^{(m)}(\xi,\tau)~\mbox{e}^{i l(kx-\omega t)},
\label{1eq:12}\\
&&\hspace*{-1.3cm}u_{i}=\sum_{m=1}^{\infty}\epsilon^{m}\sum_{l=-\infty}^{\infty}u_{il}^{(m)}(\xi,\tau)~\mbox{e}^{i l(kx-\omega t)},
\label{1eq:13}\\
&&\hspace*{-1.3cm}\phi=\sum_{m=1}^{\infty}\epsilon^{m}\sum_{l=-\infty}^{\infty}\phi_{l}^{(m)}(\xi,\tau)~\mbox{e}^{i l(kx-\omega t)},
\label{1eq:14}\
\end{eqnarray}
where $k$ and $\omega$ are real variables representing the carrier wave number and frequency,
respectively. The derivative operators can be written as
\begin{eqnarray}
&&\hspace*{-1.3cm}\frac{\partial}{\partial x}\rightarrow\frac{\partial}{\partial x}+\epsilon\frac{\partial}{\partial\xi},
\label{1eq:15}\\
&&\hspace*{-1.3cm}\frac{\partial}{\partial t}\rightarrow\frac{\partial}{\partial t}-\epsilon v_g \frac{\partial}{\partial\xi}+\epsilon^2\frac{\partial}{\partial\tau}.
\label{1eq:16}\
\end{eqnarray}
Now, by substituting Eqs. \eqref{1eq:8}-\eqref{1eq:16}  into Eqs. \eqref{1eq:1}-\eqref{1eq:4}, and \eqref{1eq:7}, and
collecting the terms containing $\epsilon$, the first order (when $m=1$ with $l=1$) reduced equations can be written as
\begin{eqnarray}
&&\hspace*{-1.3cm}n_{d1}^{(1)}=-\frac{\rho_1 k^2}{\omega^{2}}\phi_1^{(1)},
\label{1eq:17}\\
&&\hspace*{-1.3cm}u_{d1}^{(1)}=-\frac{\rho_1 k}{\omega}\phi_1^{(1)},
\label{1eq:18}\\
&&\hspace*{-1.3cm}n_{i1}^{(1)}=\frac{k^2}{{\omega^{2}-\rho_2k^2}}\phi_1^{(1)},
\label{1eq:19}\\
&&\hspace*{-1.3cm}u_{i1}^{(1)}=\frac{\omega k}{{\omega^{2}-\rho_2k^2}}\phi_1^{(1)},
\label{1eq:20}\\
&&\hspace*{-1.3cm}n_{i1}^{(1)}=(1-\rho_3)n_{d1}^{(1)}+k^2\phi_1^{(1)}+S_1\phi_1^{(1)},
\label{1eq:21}\
\end{eqnarray}
these relation provides the dispersion relation for DIAWs
\begin{eqnarray}
&&\hspace*{-1.3cm}\omega^2=\frac{M_2k^2\pm k^2\sqrt{M_2^2-4M_3M_4}}{2M_3},
\label{1eq:22}\
\end{eqnarray}
\begin{figure}[t!]
\centering
\includegraphics[width=80mm]{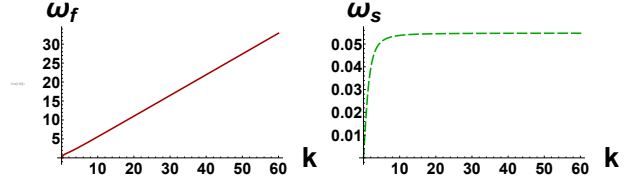}
\caption{$\omega_f$ vs $k$ (left panel) and  $\omega_s$ vs $k$ (right panel) when other plasma
parameters are $\alpha=0.5$, $q=1.6$, $\mu=2\times 10^{-6}$, $\rho=3\times 10^3$, $\rho_2=0.3$, and $\rho_3=0.5$.}
\label{1Fig:F1}
\end{figure}
where $M_2=1+\rho_1-\rho_1\rho_3+\rho_2 k^2+\rho_2 S_1$, $M_3=k^2+S_1$, and $M_4=\rho_1\rho_2-\rho_1\rho_2\rho_3$.
In Eq. \eqref{1eq:22}, to get real and positive values of $\omega$, the condition $M_2^2>4M_3M_4$ should be
satisfied. The positive and negative signs in Eq. \eqref{1eq:22} corresponds to the fast ($\omega_f$) and slow ($\omega_s$) DIA modes.
The fast DIA mode corresponds to the case in which both inertial dust and ion components oscillate in phase with the inertialess
electrons. On the other hand, the slow DIA mode corresponds to the case in which only one of the inertial components
oscillates in phase with inertialess electrons, but the other inertial component oscillates in anti-phase with
them \cite{Dubinov2009,Saberiana2017}. We have numerically analyzed the fast and slow DIA modes in Fig. \ref{1Fig:F1}
in presence of non-thermal non-extensive electrons ($\alpha=0.5$ and $q=1.6$). Figure  \ref{1Fig:F1} (left panel) indicates
that the frequency of the DIAWs can be higher than ion-plasma frequency. In this regard, we note that in absence of dust,
the frequency of the ion-acoustic waves is always less than the ion-plasma or ion-Langmuir frequency. However,
the phase speed of the DIAWs  increases with the magnitude of the dust charge ($Z_d$) and dust number
density ($n_{d0}$). This is due to the extra space charge electric field created by the highly negative
charged dust grains. This is theoretically predicted by Shukla and Silin \cite{Shukla1992} and experimentally
observed by Barkan \textit{et al.} \cite{Barkan1996}. Thus, as the magnitude of the dust charge ($Z_d$) or
dust number density ($n_{d0}$) increases, the frequency  of the DIAWs increases, even it can exceed the
ion-plasma or ion-Langmuir frequency. On the other hand, the dispersion curve of slow DIA mode shown in
Fig. \ref{1Fig:F1} (right panel) clearly indicates that the frequency of the slow DIA mode is always less
than the ion-plasma or ion-Langmuir frequency even in presence of highly negatively charged dust.
The second order equations (when $m=2$ with $l=1$) are given by
\begin{eqnarray}
&&\hspace*{-1.3cm}n_{d1}^{(2)}=-\frac{\rho_1 k^2}{\omega^{2}}\phi_1^{(2)}-\frac{2ik\rho_1(v_gk-\omega)}{\omega^{3}} \frac{\partial \phi_1^{(1)}}{\partial \xi},
\label{1eq:23}\\
&&\hspace*{-1.3cm}u_{d1}^{(2)}=-\frac{\rho_1k}{\omega}\phi_1^{(2)}-\frac{i{\rho_1}(v_gk-\omega)}{\omega^{2}}\frac{\partial \phi_1^{(1)}}{\partial \xi},
\label{1eq:24}\\
&&\hspace*{-1.3cm}n_{i1}^{(2)}=\frac{k^2}{{\omega^{2}-\rho_2k^2}}\phi_1^{(2)}
 +\frac{2ik\omega (v_gk-\omega)}{(\omega^{2}-\rho_2k^2)^2}\frac{\partial \phi_1^{(1)}}{\partial \xi},
\label{1eq:25}\\
&&\hspace*{-1.3cm}u_{i1}^{(2)}=\frac{k\omega}{{\omega^{2}-\rho_2k^2}}\phi_1^{(2)}
 +\frac{i(v_gk-\omega)(\omega^2+\rho_2k^2)}{(\omega^2-\rho_2k^2)^{2}}\frac{\partial \phi_1^{(1)}}{\partial\xi},
\label{1eq:26}\
\end{eqnarray}
with the compatibility condition
\begin{eqnarray}
&&\hspace*{-1.3cm}v_g=\frac{\rho_1\rho_3\omega^5-\rho_1\omega^5+2\rho_1\rho_2\omega^3 k^2+M_5}{\rho_1\rho_3\omega^4k-\rho_1\omega^4k+\rho_1\rho_2^2\rho_3k^5+M_6},
\label{1eq:27}\
\end{eqnarray}
where
\begin{eqnarray}
&&\hspace*{-1.3cm}M_5=-2\rho_1\rho_2\rho_3\omega^3 k^2 -\omega\rho_1\rho_2^2 k^4+\omega\rho_1\rho_3\rho_2^2 k^4-\omega^5
\nonumber\\
&&\hspace*{-0.4cm}+\omega^7-2\rho_2\omega^5 k^2+\rho_2^2\omega^3 k^4,
\nonumber\\
&&\hspace*{-1.3cm}M_6=2\rho_1\rho_2\omega^2 k^3-2\rho_1\rho_2\rho_3\omega^2 k^3-\rho_1\rho_2^2k^5-\omega^4 k.
\nonumber\
\end{eqnarray}
The coefficients of the $\epsilon$ when  $m=2$ with $l=2$ provides the second
order harmonic amplitudes which are found to be proportional
to $|\phi_1^{(1)}|^2$
\begin{eqnarray}
&&\hspace*{-1.3cm}n_{d2}^{(2)}=S_4|\phi_1^{(1)}|^2,
\label{1eq:28}\\
&&\hspace*{-1.3cm}u_{d2}^{(2)}=S_5 |\phi_1^{(1)}|^2,
\label{1eq:29}\\
&&\hspace*{-1.3cm}n_{i2}^{(2)}=S_6|\phi_1^{(1)}|^2,
\label{1eq:30}\\
&&\hspace*{-1.3cm}u_{i2}^{(2)}=S_7 |\phi_1^{(1)}|^2,
\label{1eq:31}\\
&&\hspace*{-1.3cm}\phi_{2}^{(2)}=S_8|\phi_1^{(1)}|^2,
\label{1eq:32}\
\end{eqnarray}
where
\begin{eqnarray}
&&\hspace*{-1.3cm}S_4=\frac{3\rho_1^2k^4-2\rho_1\omega^2k^2S_8}{2\omega^4},
\nonumber\\
&&\hspace*{-1.3cm}S_5=\frac{\rho_1^2k^3-2\rho_1\omega^2kS_8}{2\omega^3},
\nonumber\\
&&\hspace*{-1.3cm}S_6=\frac{2k^2S_8(\omega^2-\rho_2k^2)^2+k^4(\rho_2 k^2+3\omega^2)}{2(\omega^2-\rho_2k^2)^3},
\nonumber\\
&&\hspace*{-1.3cm}S_7=\frac{\omega S_6(\omega^2-\rho_2k^2)-\omega k^4}{k(\omega^2-\rho_2k^2)^2},
\nonumber\\
&&\hspace*{-1.3cm}S_8=\frac{k^4\omega^4(\rho_2 k^2+3\omega^2)-(3\rho_1^2k^4+2\omega^4S_2)(\omega^2-\rho_2 k^2)^3}{6k^2\omega^4(\omega^2-\rho_2 k^2)^3}.
\nonumber\
\end{eqnarray}
When $m=3$ with $l=0$ and $m=2$ with $l=0$ lead to zeroth harmonic modes as follows
\begin{eqnarray}
&&\hspace*{-1.3cm}n_{d0}^{(2)}=S_{9}|\phi_1^{(1)}|^2,
\label{1eq:33}\\
&&\hspace*{-1.3cm}u_{d0}^{(2)}=S_{10}|\phi_1^{(1)}|^2,
\label{1eq:34}\\
&&\hspace*{-1.3cm}n_{i0}^{(2)}=S_{11}\phi_1^{(1)}|^2,
\label{1eq:35}\\
&&\hspace*{-1.3cm}u_{i0}^{(2)}=S_{12}|\phi_1^{(1)}|^2,
\label{1eq:36}\\
&&\hspace*{-1.3cm}\phi_0^{(2)}=S_{13} |\phi_1^{(1)}|^2,
\label{1eq:37}\
\end{eqnarray}
where
\begin{eqnarray}
&&\hspace*{-1.3cm}S_{9}=\frac{\rho_1^2k^2\omega+2v_g\rho_1^2k^3-\rho_1S_{13}\omega^3}{v_g^2\omega^3},
\nonumber\\
&&\hspace*{-1.3cm}S_{10}=\frac{k^2\rho_1^2-\rho_1S_{13}\omega^2}{v_g\omega^2},
\nonumber\\
&&\hspace*{-1.3cm}S_{11}=-\frac{{S_{13}(\omega^2-\rho_2k^2)^2+k^2(\rho_2k^2+\omega^2+2\omega v_g k)}}{(\rho_2-v_g^2)(\omega^2-\rho_2k^2)^2},
\nonumber\\
&&\hspace*{-1.3cm}S_{12}=\frac{v_gS_{11}(\omega^2-\rho_2k^2)^2-2\omega k^3} {(\omega^2-\rho_2k^2)^2},
\nonumber\\
&&\hspace*{-1.3cm}S_{13}=\frac{2S_2v_g^2\omega^3(\rho_2-v_g^2)(\omega^2-\rho_2k^2)^2+M_7}{\omega^3(\omega^2-\rho_2k^2)^2\times M_8},
\nonumber\
\end{eqnarray}
with
\begin{eqnarray}
&&\hspace*{-1.3cm}M_7=k^2v_g^2\omega^3(\rho_2k^2+\omega^2)+2k^3v_g^3\omega^4
\nonumber\\
&&\hspace*{-0.5cm}+\rho_1^2k^2\omega(1-\rho_3)(\rho_2-v_g^2)(\omega^2-\rho_2k^2)
\nonumber\\
&&\hspace*{-0.5cm}+2v_g\rho_1^2 k^3(1-\rho_3)(\rho_2-v_g^2)(\omega^2-\rho_2k^2),
\nonumber\\
&&\hspace*{-1.3cm}M_8=\rho_1(1-\rho_3)(\rho_2-v_g^2)-v_g^2-v_g^2S_1(\rho_2-v_g^2).
\nonumber\
\end{eqnarray}
Finally, the third harmonic modes when $m=3$ and $l=1$ with the help of \eqref{1eq:17}-\eqref{1eq:37}, give a
set of equations, which can be reduced to the standard NLSE:
\begin{eqnarray}
&&\hspace*{-1.3cm}i\frac{\partial\Phi}{\partial\tau}+P\frac{\partial^2\Phi}{\partial \xi^2}+Q|\Phi|^2\Phi=0,
\label{1eq:38}\
\end{eqnarray}
where $\Phi=\phi_1^{(1)}$ for simplicity. In equation \eqref{1eq:38}, dispersion
coefficient $P$  can be written as
\begin{eqnarray}
&&\hspace*{-1.3cm}P=-\frac{2kv_g\omega^5(\omega v_g-\rho_2 k)(\omega-v_g k)+M_{9}}{\omega (\omega^2-\rho_2 k^2)\times M_{10}},
\nonumber\
\end{eqnarray}
where
\begin{eqnarray}
&&\hspace*{-1.3cm}M_9=\rho_1(1-\rho_3)(\omega-v_g k)(4v_gk-\omega)(\omega^2-\rho_2 k^2)^3
\nonumber\\
&&\hspace*{-0.5cm}+\omega^4(\omega^2-\rho_2 k^2)^3-\omega^4(\omega-v_g k)^2(\omega^2+\rho_2 k^2),
\nonumber\\
&&\hspace*{-1.3cm} M_{10}=2\omega^4 k^2+2\rho_1 k^2(1-\rho_3)(\omega^2-\rho_2 k^2)^2,
\nonumber\
\end{eqnarray}
and also  the nonlinear coefficient $Q$  can be written as
\begin{eqnarray}
&&\hspace*{-1.3cm}Q=\frac{M_{11}}{2\omega^4 k^2+2\rho_1 k^2(1-\rho_3)(\omega^2-\rho_2 k^2)^2},
\nonumber\
\end{eqnarray}
where
\begin{eqnarray}
&&\hspace*{-1.3cm}M_{11}=3\omega^3(\omega^2-\rho_2 k^2)^2S_3
\nonumber\\
&&\hspace*{-0.4cm}+2\omega^3(\omega^2-\rho_2 k^2)^2S_2(S_8+S_{13})
\nonumber\\
&&\hspace*{-0.4cm}-\omega\rho_1 k^2(1-\rho_3)(S_4+S_9)(\omega^2-\rho_2 k^2)^2
\nonumber\\
&&\hspace*{-0.4cm}-2\rho_1 k^3(1-\rho_3)(S_5+S_{10})(\omega^2-\rho_2 k^2)^2
\nonumber\\
&&\hspace*{-0.4cm}-2\omega^4 k^3(S_7+S_{12})-\omega^3(\omega^2k^2+\rho_2k^4)(S_6+S_{11}).
\nonumber\
\end{eqnarray}
\begin{figure}[t!]
\centering
\includegraphics[width=80mm]{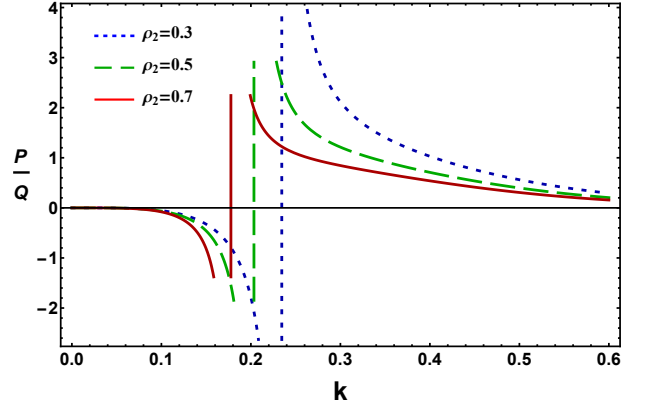}
\caption{The variation of $P/Q$ with $k$  for different values of $\rho_2$
when other plasma parameters are $\alpha=0.5$, $q=1.6$, $\mu=2\times 10^{-6}$, $\rho=3\times 10^3$, $\rho_3=0.5$, and $\omega_f$.}
\label{1Fig:F2}
\end{figure}
It is interesting that the dispersive coefficient $P$ and nonlinear coefficient $Q$
of the NLSE \eqref{1eq:38} are function of various plasma parameters such as carrier wave number $k$, the ratio of positive
ion mass to negative dust mass (via $\mu$), the ratio of negative dust charge state to positive ion
charge state (via $\rho$), ion temperature to electron temperature times the charge state of the positive
ion (via $\rho_2$), the number density of electron to the number density of positive ion times
the charge state of the positive ion (via $\rho_3$),  the non-thermality (via $\alpha$) and
non-extensivity (via $q$) properties of electrons, etc.
\section{MI and Rogue Waves}
\label{1sec:Modulational Instability and Rogue Waves}
To study the MI of DIAWs, we consider the linear solution of the
NLSE \eqref{1eq:38} in the form $\Phi=\widetilde{\Phi}e^{iQ|\widetilde{\Phi}|^2\tau}$+c.c.,
where $\widetilde{\Phi}=\widetilde{\Phi}_0+\epsilon\widetilde{\Phi}_1$ and
$\widetilde{\Phi}_1=\widetilde{\Phi}_{1,0}e^{i(\widetilde{k}\xi-\widetilde{\omega}{ \tau})}+c.c$.
We note that the amplitude depends on the frequency, and that the perturbed wave number $\widetilde{k}$
and frequency $\widetilde{\omega}$ which are different from $k$ and $\omega$. Now, substituting these
into NLSE \eqref{1eq:38}, one can easily obtain the following nonlinear
dispersion relation \cite{Kourakis2003,Fedele2002,Sultana2011,Bains2013,Rahman2018}
\begin{eqnarray}
&&\hspace*{-1.3cm}\widetilde{\omega}^2=P^2\widetilde{k}^2\Big(\widetilde{k}^2-\frac{2|\widetilde{\Phi}_0|^2}{P/Q}\Big).
\label{1eq:39}
\end{eqnarray}
It is observed here that the ratio $P/Q$ is negative (i.e., $P/Q<0$), the DIAWs
will be modulationally stable. On the other hand, if the ratio $P/Q$ is positive
(i.e., $P/Q>0$), the DIAWs will be modulationally unstable. It is obvious from Eq. \eqref{1eq:39}
that the DIAWs becomes modulationally unstable when $\widetilde{k}_c>\widetilde{k}$ in
the regime $P/Q>0$, where $\widetilde{k}_c = \sqrt{2(Q/P)}{|\widetilde{\Phi}_0|}$.
The growth rate $\Gamma$ of the modulationally unstable DIAWs is given by
\begin{eqnarray}
&&\hspace*{-1.3cm} \Gamma=|P|\widetilde{k}^2\sqrt{\frac{\widetilde{k}_c^2}{\widetilde{k}^2}-1}.
\label{1eq:40}
\end{eqnarray}
\begin{figure}[t!]
\centering
\includegraphics[width=80mm]{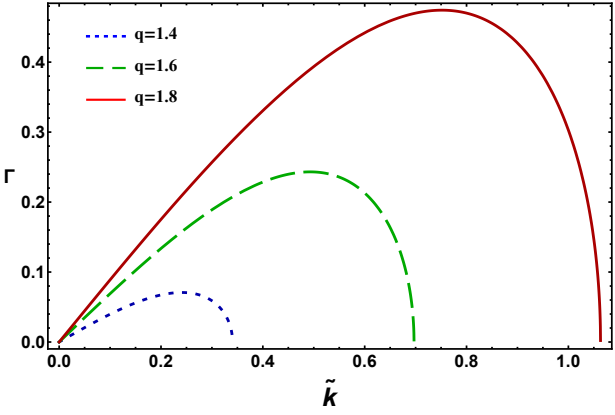}
\caption{The variation of $\Gamma$ with $\widetilde{k}$ for different values of $q$
when other plasma parameters are $\alpha=0.5$, $\mu=2\times 10^{-6}$, $\rho=3\times 10^3$, $\rho_2=0.3$, $\rho_3=0.5$, $k=0.4$, $\Phi_0=0.5$, and $\omega_f$.}
\label{1Fig:F3}
\end{figure}
\begin{figure}[t!]
\centering
\includegraphics[width=80mm]{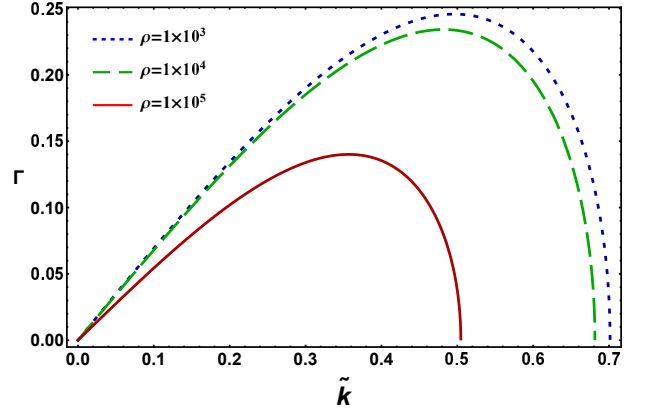}
\caption{The variation of $\Gamma$ with $\widetilde{k}$ for different values of $\rho$
when other plasma parameters are  $\alpha=0.5$, $q=1.6$, $\mu=2\times 10^{-6}$, $\rho_2=0.3$, $\rho_3=0.5$, $k=0.4$, $\Phi_0=0.5$, and $\omega_f$.}
\label{1Fig:F4}
\end{figure}
Equation \eqref{1eq:40} can easily express the nonlinear character of the DPM.
The first-order rational solution, in the domain where dispersive and nonlinear coefficients of NLSE have an opposite
sign (i.e., $P/Q>0$), can be written as  \cite{Akhmediev2009}
\begin{eqnarray}
&&\hspace*{-1.3cm}\Phi(\xi,\tau)= \sqrt{\frac{2P}{Q}} \Big[\frac{4(1+4iP\tau)}{1+16P^2\tau^2+4\xi^2}-1 \Big]e^{2iP\tau},
\label{1eq:41}\
\end{eqnarray}
\begin{figure}[t!]
\centering
\includegraphics[width=80mm]{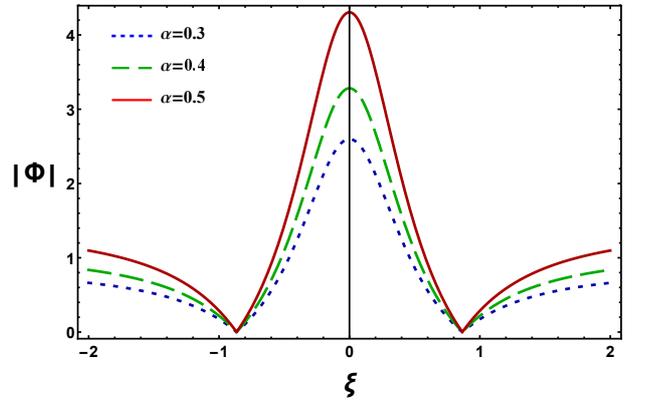}
\caption{The variation of $|\Phi|$ with $\xi$  for different values of $\alpha$ when other plasma parameters are $q=1.6$,
$\mu=2\times 10^{-6}$, $\rho=3\times 10^3$, $\rho_2=0.3$, $\rho_3=0.5$, $k=0.4$, $\tau=0$, and $\omega_f$.}
\label{1Fig:F5}
\end{figure}
\begin{figure}[t!]
\centering
\includegraphics[width=80mm]{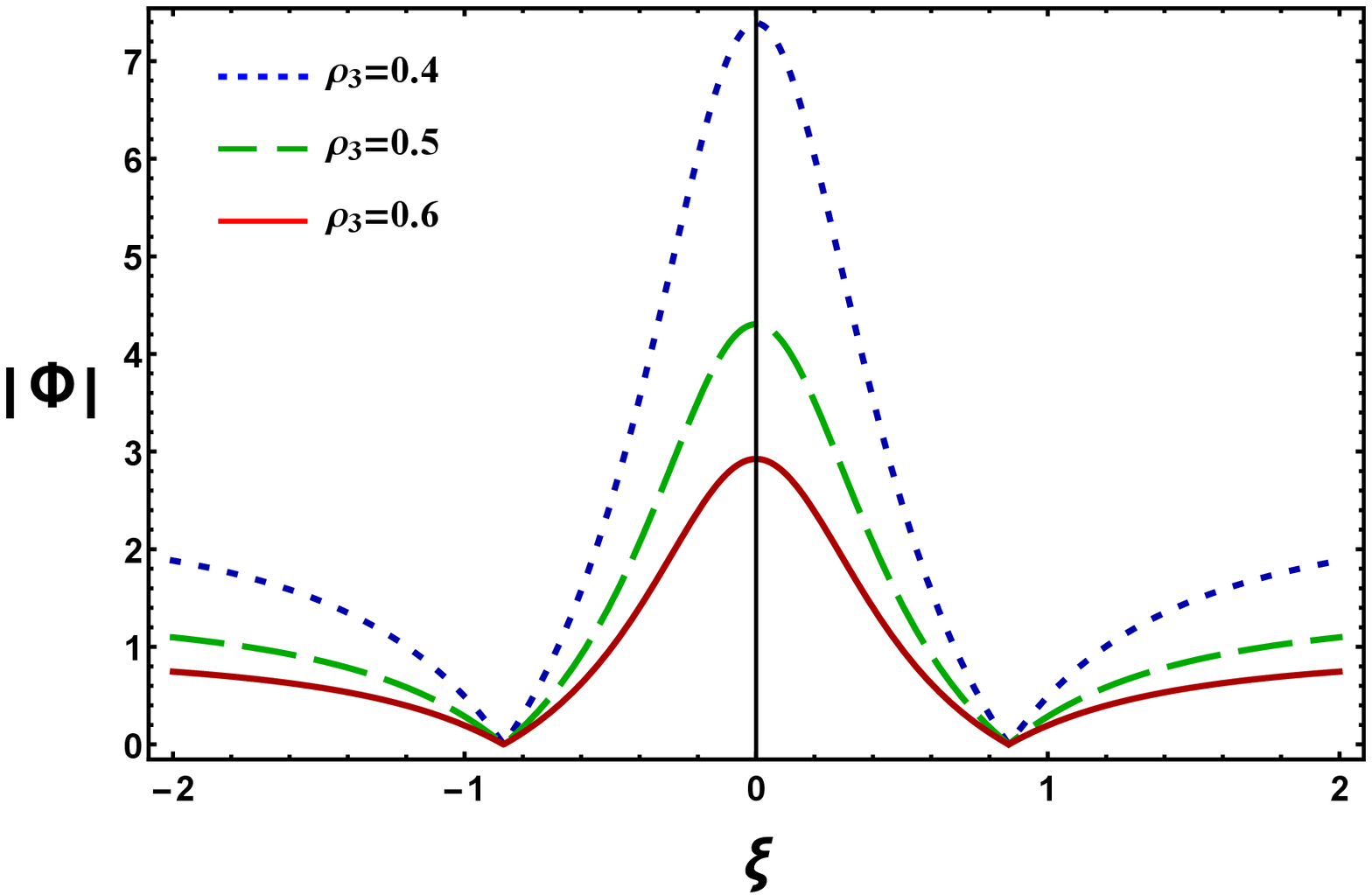}
\caption{The variation of $|\Phi|$ with $\xi$  for different values of $\rho_3$ when other plasma parameters are $\alpha=0.5$, $q=1.6$,
$\mu=2\times 10^{-6}$, $\rho=3\times 10^3$, $\rho_2=0.3$, $k=0.4$, $\tau=0$, and $\omega_f$.}
\label{1Fig:F6}
\end{figure}
\begin{figure}[t!]
\centering
\includegraphics[width=80mm]{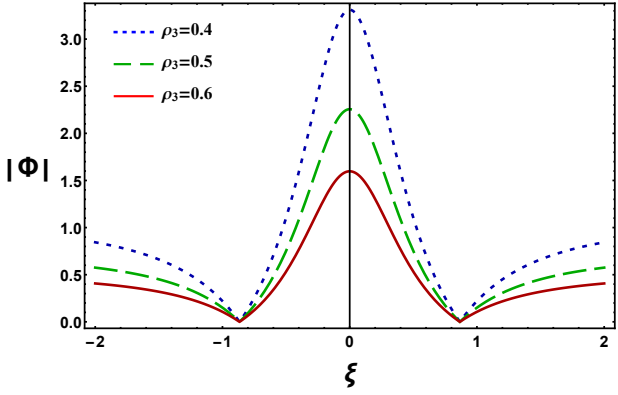}
\caption{The variation of $|\Phi|$ with $\xi$  for different values of $\rho_3$ when other plasma parameters are $\alpha=0$, $q=1$,
$\mu=2\times 10^{-6}$, $\rho=3\times 10^3$, $\rho_2=0.3$, $k=0.4$, $\tau=0$, and $\omega_f$.}
\label{1Fig:F7}
\end{figure}
The Eq. \eqref{1eq:41} implies that the concentration of high energy occurs within a small region.
\section{Numerical analysis}
\label{1sec:Numerical analysis}
Now, we would like to numerically analyze the stability conditions of the DIAWs in 
presence of the non-thermal non-extensive  electrons. The mass and charge state of 
the plasma species even their number density are important factors in recognizing the 
stability conditions of the DIAWs in DPM \cite{Saberiana2017,Merlino2014,Shukla2012,Mamun2002,Shalaby2009,Shukla2002}.
The mass of the dust grains is comparable to the mass of proton. In general picture
of the DPM, dust grains are massive (millions to billions times heavier than the
protons) and their sizes range from nanometres to millimetres. Dust grains may be metallic, conducting, or made of ice
particulate. The size and shape of dust grains will be different, unless they are man-made.
The dust grains are millions to billions times heavier than the protons, and typically, a dust grain acquires one
thousand to several hundred thousand elementary charges \cite{Saberiana2017,Merlino2014,Shukla2012,Mamun2002,Shalaby2009,Shukla2002}.
In this article, we consider three components dusty plasma model having inertial warm positive ions and negative dust grains,
and interialess non-thermal non-extensive electrons. It may be noted here that in the DIAWs, if anyone consider
the pressure term of the ions then it is important to be considered the moment of inertia of the ions along with
the dust grains in presence of inertialess electrons. This means that the consideration of the pressure term
of the ions highly contributes to the moment of inertia along with inertial dust grains to generate DIAWs in a DPM having inertialess electrons.
In our present analysis, we have considered that $m_d=10^6m_i$, $Z_d=(10^3\thicksim10^5)Z_i$, and $T_e=10T_i$.

The variation of $P/Q$ with $k$ for different values of $\rho_2$ is shown in Fig. \ref{1Fig:F2}
which provides the information about the effects of the temperature of the positive ion ($T_i$)
and negative electron ($T_e$), and  the charge state of the positive ion $Z_{i}$ in presence of
non-Maxwellian (i.e., $\alpha=0.5$ and $q=1.6$) electrons. The point, in which $P/Q$ curve coincides
with the $k$-axis in $P/Q$ versus $k$ graph, is known as threshold/critical wave number ($k_c$) which
divides the unstable domain from stable one for the DIAWs. It is clear from this figure that when
$\rho_2=0.3$, $0.5$, $0.7$, then the corresponding value of $k_c\cong0.24$ (dotted blue curve), $0.20$
(dashed green), $0.18$ (solid red curve). Both modulationally stable (i.e., for long wavelength)
and unstable (i.e., for short wavelength) domains of DIAWs are allowed by non-thermal non-extensive DPM,
and if we increase (decrease) in the value of $\rho_2$ which can shift the $k_c$ to a lower (higher) value of $k$
and this means that the DIAWs will be unstable for small values of $k$ with an increase in the ion temperature while
stable for the large values of $k$  with an increase in the electron temperature.

We have numerically analyzed Eq. \eqref{1eq:40} in  Figs. \ref{1Fig:F3}
and \ref{1Fig:F4} by using these plasma parameters: $\alpha=0.5$, $\mu=2\times 10^{-6}$,
$\rho_2=0.3$, $\rho_3=0.5$, $\Phi_0=0.5$, $k=0.4$, $\omega_f$, and have also maintained the
assumption $m_d=10^6m_i$, $Z_d=(10^3\thicksim10^5)Z_i$, and $T_e=10T_i$ for the depiction of
the growth rate of DIAWs. The variation of the $\Gamma$ with $\widetilde{k}$ indicates that initially, the
$\Gamma$ increases with $\widetilde{k}$ and becomes maximum for a particular value of $\widetilde{k}$
then again decreases to zero. It can be seen from Fig. \ref{1Fig:F3} that the maximum value of the growth rate increases with
the increase in the value of the $q$, and this result is a  good agreement with the result
of Ref. \cite{Rahman2018}. The effects of the charge state of the negative dust ($Z_d$) and
positive ion ($Z_i$) on the growth rate of the MI of DIAWs are displayed in Fig. \ref{1Fig:F4}
which indicates that the $\Gamma$ decreases (increases) with $Z_d$ ($Z_i$).
Physically, the nonlinearity of the plasma medium as well as maximum value
of the $\Gamma$ increases with $Z_i$ while the nonlinearity of the plasma medium as well as maximum value
of the $\Gamma$ decreases with $Z_d$.

The effects of the non-thermality of electrons  (via $\alpha$) on the amplitude and width of the
DIARWs, which can be depicted by using Eq. \eqref{1eq:41}, can be seen from
Fig. \ref{1Fig:F5}. This figure clearly indicates
that the increase in the value of $\alpha$ does not only cause to increase the amplitude
of the DIARWs associated DIAWs in the modulationally unstable domain (i.e., $P/Q>0$)
but also cause to increase the width of the DIARWs associated DIAWs in the
modulationally unstable domain (i.e., $P/Q > 0$). The physics of this result is that increasing
$\alpha$ causes to increase the nonlinearity of the plasma medium as well as the amplitude and
width of the DIARWs, and this result is a nice agreement with the result of
El-Labany \textit{et al.} \cite{El-Labany2015} and Hassan \textit{et al.} \cite{Hassan2019}.

The nature of the DIARWs, which can be depicted by using Eq. \eqref{1eq:41},
with the variation of $\rho_3$ can be observed from Figs. \ref{1Fig:F6} and \ref{1Fig:F7}. It is easy to demonstrate
from these two figures that increasing (decreasing) the values of $\rho_3$ leads to a decrease (increase)
of the amplitude and width of the DIARWs, which indicates that an increase (decrease) of $\rho_3$
could  shrink (enhance) the amplitude and width of the DIARWs.
Actually, the nonlinearity as well as the amplitude and width of the DIARWs associated with DIAWs
in the modulationally unstable domain (i.e., $P/Q > 0$) increase as we increase (decrease) the
value of $n_{i0}$  ($n_{e0}$) for a fixed value of $Z_i$.
It is interesting from Figs. \ref{1Fig:F6} and \ref{1Fig:F7} that the order of the
variation of DIARWs amplitude and width is not depend on the presence of the Maxwellian/non-Maxwellian electrons, and
in presence of non-Maxwellian electrons (i.e., $\alpha=0.5$ and $q=1.6$), the amplitude of the DIARWs is always greater
than the presence of Maxwellian electrons  (i.e., $\alpha=0$ and $q=1$) for same value of $\rho_3$.

Finally, under consideration (i.e., $P/Q>0$), the DPM  supports the formation of DIARWs associated
with DIAWs, and we have already graphically observed the variation of the DIARWs with respect to $\xi$ at $\tau=0$ in
Figs. \ref{1Fig:F5} to \ref{1Fig:F7}, and these figures clearly describe that the  RWs
 has two minima and one maxima in the potential which is the typical
feature of the DIARWs. This property reveals that our plasma system can concentrate a
significant amount of DIA wave energy in a relatively small region at $\tau=0$.
\section{Conclusion}
\label{1sec:Conclusion}
To conclude, we have considered a DPM having inertial warm positive ions along with inertial
negative dust grains, and inertialess electrons featuring non-thermal non-extensive
distribution. Two types of DIA modes can be found from the theoretical calculation, and
the frequency  of the DIAWs increases as the magnitude of the dust charge or dust number density increases
even it can exceed the ion-plasma or ion-Languir frequency. We have also numerically observed the
characterizes of the MI growth rate of the DIAWs and it is found that the maximum value of the growth rate
increases with the increase in the value of the non-extensive parameter and the charge state of the
positive ion while decreases with the charge state of negative dust.
We have also numerically examined the variation of rational solutions of the NLSE \eqref{1eq:38}
with the plasma parameters $\alpha$ and $\rho_3$. It is found
that the amplitude and width of the DIARWs increases with $\alpha$ but decreases with $\rho_3$.
It is interesting fact that in presence of non-Maxwellian electrons, the amplitude of the DIARWs is always greater
than the presence of Maxwellian electrons  for same value of $\rho_3$. It is noted here that the simulation
of our present investigation is very important for this new kind DIAWs but  beyond the
scope of our present work. We, finally, propose that the findings of our present investigation should be useful
to understand the nonlinear phenomena (viz. the MI of DIAWs and formation of DIARWs) in a DPM 
where electrons follow non-extensive non-thermal distribution.

\end{document}